\documentclass[conference]{IEEEtran}
\IEEEoverridecommandlockouts
\usepackage{cite}
\usepackage{amsmath,amssymb,amsfonts}
\usepackage{algorithmic}
\usepackage{graphicx}
\usepackage{textcomp}
\usepackage{adjustbox}
\usepackage{xcolor}
\usepackage{hyperref} 
\usepackage{setspace}

\def\BibTeX{{\rm B\kern-.05em{\sc i\kern-.025em b}\kern-.08em
    T\kern-.1667em\lower.7ex\hbox{E}\kern-.125emX}}

\makeatletter
\def\@IEEEpubidpullup{8\baselineskip}
\makeatother

\begin{document}

\title{Deep Personalized Re-targeting\\
}

\author{
\IEEEauthorblockA{}
}

\author{\IEEEauthorblockN{\IEEEauthorblockN{Meisam Hejazinia\IEEEauthorrefmark{1},
Pavlos Mitsoulis-Ntompos\IEEEauthorrefmark{2}, Serena Zhang\IEEEauthorrefmark{3}}
}
\IEEEauthorblockA{Vrbo, part of Expedia Group\\
\IEEEauthorrefmark{1}mnia@expediagroup.com,
\IEEEauthorrefmark{2}pntompos@expediagroup.com,
\IEEEauthorrefmark{3}shuazhang@expediagroup.com
}}



\maketitle

\begin{abstract}
Predicting booking probability and value at the traveler level plays a central role in computational advertising for massive two-sided vacation rental marketplaces. These marketplaces host millions of travelers with long shopping cycles, spending a lot of time in the discovery phase. The footprint of the travelers in their discovery is a useful data source to help these marketplaces to predict shopping probability and value. However, there is no one-size-fits-all solution for this purpose. In this paper, we propose a hybrid model that infuses deep and shallow neural network embeddings into a gradient boosting tree model. This approach allows the latent preferences of millions of travelers to be automatically learned from sparse session logs. In addition, we present the architecture that we deployed into our production system. We find that there is a pragmatic sweet spot between expensive complex deep neural networks and simple shallow neural networks that can increase the prediction performance of a model by seven percent, based on offline analysis.
\end{abstract}

\begin{IEEEkeywords}
computational advertising, re-targeting, personalized advertising, shopping funnel, deep learning, embeddings, e-commerce, gradient boosting
\end{IEEEkeywords}

\section{Introduction}
Every day, millions of travelers enter massive online two-sided marketplaces from various advertising channels such as search engines, display ads and meta search engines to discover and book their dream vacation property from millions of options. A significant portion of these travelers have never booked a vacation rental before, or are not willing to repeat their previous trips as they seek variety. Marketplace platforms must bid appropriately in order to gain traffic from these travelers through online advertising channels. To accomplish this, marketplaces may estimate the booking probability and value of potential travelers based on various engagement signals. While traditional customer lifetime value estimation methods are strong in the context of repeated purchases, these methods fall short in the context of infrequent or first time travelers. This paper attempts to fill in this gap by proposing a solution that leverages search and engagement signals to predict booking intent and value for a traveler as they progress through the shopping cycle, regardless of their previous booking history. In particular, we propose a hybrid method that infuses shallow and deep neural network embeddings into a gradient boosting tree model in order to automatically extract features that improve booking intent and value prediction. We discover a pragmatic sweet spot between expensive complex deep neural networks and shallow neural networks that improves the shopping intent prediction model by seven percent. We explain how our method is lightweight in terms of computational resources and easily deployable into a large-scale production system.  

There are four areas of study related to this paper including probabilistic customer lifetime value, machine learning in computational advertising and click through rate (CTR) prediction, embeddings, and deep neural networks.  We compare and contrast this study with relevant studies in each of the areas in the following subsections. 
\subsection{Probabilistic Customer Lifetime Value}
Historically, businesses have leveraged approaches to value their customers in order to optimize their return on advertising spend (ROAS) \cite{wang2017display}, particularly for traditional media, affiliate marketing, display ads, and search engine advertising. Industry has adopted probabilistic and statistical learning approaches due to their explainability, availability of small amounts of data, and computational restrictions. These models put simplified assumptions on the time between purchase, time til churn event, purchase count, and purchase value, proposing the following structures: Bayesian Pareto/ Negative Binomial Distribution baseline (NBD), Beta-Gamma(BG) /NBD, Gamma/Gompertz, and various enhancement on them \cite{fader2005counting}  \cite{fader2016pareto} \cite{fader2009probability} \cite{glady2009modified} \cite{chan2011measuring}. As computational power allowed, more complex approaches in the same vein as Bayesian Hierarchical Hidden Markov Model (HHMM) were proposed \cite{netzer2008hidden}. While these studies open up a path for valuing customers, they might not be the right approach for re-targeting in the massive online marketplace with millions of travelers and listings, where travelers are heterogeneous, seeking variety in experience, not identifying themselves, and booking infrequently. In such an environment, we need approaches that are not only scalable, but also leverage various pre-purchase signals,  such as browsing behavior to predict shopping intent and value. Furthermore, traveler's preference shifts from trip to trip, making the historical behavior less relevant. Machine learning approaches have a relatively strong track record handling such cases. 
\subsection{Machine learning in computational advertising}
In more recent years, industry practitioners leveraged machine learning approaches for re-targeting at scale. Due to their ability to generate superior predictions, tree based approaches, such as random forest and gradient boosted trees were first to gain popularity  \cite{juan2016field} \cite{chen2016xgboost} \cite{vanderveld2016engagement}. However, these approaches required a significant effort in feature engineering, making them hard to generalize and expensive to maintain. To overcome this limitation, many studies began to leverage deep neural networks to automatically generate latent features in this domain \cite{10.1007/978-3-030-10997-4_9} \cite{Sheil2018PredictingPI} \cite{wang2017deep} \cite{arava2018deep} \cite{gai2017learning} \cite{zhou2018deep}. While these approaches have achieved significant improvements, they require a lot of time to train on a multi-million by multi-million sparse-space of traveler and listing graph.  There might be a sweet spot between the tree based methods and deep neural networks which under computational budget constraints delivers better results. Next, we review embedding methods which are reputed for handling sparse spaces, and then we review the history of deep neural networks to introduce the space we explore in this study.
\subsection{Embedding methods}
The data sparsity problem has long been studied in the recommendation system literature. Collaborative filtering methods used projection or embeddings of user-item matrix in lower dimensional space through matrix decomposition \cite{guigoures2018hierarchical}  \cite{wan2015next}  \cite{chang2018content}  \cite{liang2016factorization}  \cite{johnson2014logistic}  \cite{barkan2016item2vec}. More recently session data has been proposed as an alternative to purchase data for collaborative filtering methods, generating a new trend in recommendation system literature that focuses on session based recommendation systems (SBRS)  \cite{wang2019survey} \cite{wu2018session}. In essence, the methods used in SBRS extend natural language processing (NLP) methods to predict the context of user browsing, defined by the items the user views before or after a given item, using neural network methods \cite{mikolov2013distributed}. The advantage of these methods is that they can project millions of items into a lower dimensional space, in which contextually similar items appear close to each-other \cite{grbovic2015commerce} \cite{barkan2016item2vec} \cite{grbovic2018real} \cite{wang2018billion} \cite{arora2016simple} \cite{vasile2016meta} \cite{caselles2018word2vec} \cite{bogina2017incorporating}.  These lower dimension representations can be leveraged as automatically generated features in shopping intent and valuation models conditional on the user activity \cite{shan2016deep} \cite{cheng2016wide}.  Many of these studies extend the embedding space from shallow to deep neural network models. 

\subsection{Deep Neural Network}
Due to their success in text and image domain, deep neural networks have recently gained popularity in recommendation systems and embedding spaces \cite{iyyer2015deep} \cite{sedhain2015autorec} \cite{zhu2018learning}. Recurrent neural networks have shown to be effective in capturing temporal dependencies between user item views \cite{lang2017understanding} \cite{baral2018close} \cite{eide2018deep} \cite{iyyer2015deep} \cite{bogina2017incorporating} and convolutional neural networks have shown success in capturing latent intent structure in the item images \cite{prevost2018deep} \cite{eide2018deep} \cite{liu2019feature}. Recently, a new architecture called attention networks have gained popularity, due to its ability to automatically reweigh all signals that users can capture, resembling user memory attention \cite{chaudhari2019attentive} \cite{1904.05985} \cite{zhou2016attention} \cite{lake2019large}. In this study, we evaluate the merits of various candidate solutions such as Deep Average Networks, Long Short Term Memory, and Attention networks in adding value to shopping intent and value prediction.
The closest study to this paper is \cite{chamberlain2017customer}, which suggests using embeddings in customer lifetime value prediction. However, our paper differentiates itself by extending the embedding features from linear to non-linear spaces, finding a sweet spot with low cost and high value in combining a simple deep neural network with a shallow neural network and a tree based method to predict shopping intent and value.

\section{Notations and Problem Formulation}

We formalize the personalized re-targeting problem as follows. The traveler performs a set of activities $\mathrm{S}_j = \left\{a_1, a_2,..., a_T\right\}$ in a traveler shopping session $\mathrm{S}_j$ when they visit our two-sided platform, where $T$ is the length of the sequence. An activity here could be either a click or page view. We also represent each session $\mathrm{S}_j$ as a set of listings that the traveler interacts with $\left\{l_1, l_2,..., l_T\right\}$, where $l_t \in \mathbb{L}$. At the end of multiple visit sessions the traveler may either make a booking, represented by $Y_j = 1$ or they might leave the platform for another time, represented by $Y_j=0$. We denote probability of booking conditional on historical session context $C_j$ with $\mathrm{P}(Y_j|\mathrm{S}_j, C_j)$,  and conditional probability of booking value $\mathrm{P}(V_j|\mathrm{S}_j, C_j)$. For re-targeting returning travelers, we want to estimate the distribution of both booking event and booking value.

In an advertising real time bidding (RTB) system (such as Bing Ads, Google AdWords, and Criteo), a quantitative bidding function for bid utility $\mathcal{U}$ can be boiled into two components: the estimated utility of the ad opportunity and the estimated cost to win it \cite{wang2017display}. The hypothesis is that all other information and the bid price are conditionally independent given these two components. Further, we can decompose the utility of bidding $\mathcal{U}$ into two components: first the on-site conversion  $\mathcal{R}$, and second marginal value $\mathcal{M}$, in summary $\mathcal{U} = \mathcal{R} * \mathcal{M}$. Ideally, we should optimize this objective function under budget constraints, yet we focus on estimating the utility in this paper. In particular, our methodology estimates $\mathrm{P}(Y_j|\mathrm{S}_j, C_j)$ and $\mathrm{P}(V_j|\mathrm{S}_j, C_j)$, to extract $\mathcal{R}$ and $\mathcal{M}$, respectively, to re-target travelers for their next visit in the shopping funnel. We will not cover other aspects of the bid optimization here. We propose a hybrid deep learning framework deployed in an end-to-end fashion in order to solve the above challenges. 

\section{Model}
In this section, we describe our solutions to the personalized re-targeting problem. Our solution has two modular parts: the conversion prediction and marginal value prediction. We use the XGBoost algorithm in both of the components. Later, we will focus on describing our ongoing efforts to supplement the handcrafted features in the deployed systems with automatic feature learning using traveler embeddings. The traveler embeddings are generated based on a two-stage neural network model described below.  

\subsection{Traveler Booking Intent and Booking Value XGBoost Models}
We used XGBoost, a scalable machine learning system for gradient tree boosting \cite{friedman2000additive} \cite{chen2016xgboost} in our solutions for both modular parts. There are a few advantages of using XGBoost for this problem:
\begin{itemize}
    \item Tree boosting has been shown to give state-of-the-art results on many standard  classification benchmarks.
    \item Parallel and distributed computation and capability to handle sparse data.
    \item Deployment of XGBoost in an end-to-end system efficiently scales to large data sets.
\end{itemize}

Feature engineering and domain dependent data analysis play a pivotal role in this solution. For this model, we have created a rich set of handcrafted features using session data that includes user onsite interactions, e.g. search, property listing page, check out flow and contacts (inquiries and messages). An in-house model was used to remove bot traffic. The session data are further aggregated cross the entire period for each traveler $j$ to obtain the final set of features and labels $\mathcal{D} = \left\{(\textbf{x}_j, y_j) \right\}$ ($\textbf{x}_j \in \mathbb{R}^n$, $y_j \in \mathbb{R}$), where $y_j$ is the binary booking indicator in the booking intent model and continuous booking value in the booking value model.

Within the XGBoost framework, the objective is using additive functions to predict the output $\hat{y_j} \equiv  F(\mathbf{x}_j) = \sum_{k=1}^K f_k(x_i)$, where $f_k \in \mathbb{F}$ and $\mathbb{F}$ the space of classification and regression trees. The following regularized objective function is optimized in an incremental and greedy fashion:
\begin{equation}\label{eq:1}
    {L}^{(t)} = \sum_{j} l(y_j, \hat{y}_j^{(t-1)} + f_t(\mathbf{x}_j)) + \Omega(f_t)
\end{equation}
where $\hat{y}_j^{(t)}$ is the prediction of the $j$-th instance at the $t$-th iteration, $l$ is the loss function (log likelihood) that measures the difference between the prediction and the target, and $\Omega(f_t)$ is a regularization term which penalizes the complexity of the tree functions. XGBoost uses second order approximation to fit a base learner $f_t(\mathbf{x}_j)$ to minimize (\ref{eq:1}) at each iteration. In addition, shrinkage and column sub-sampling are employed in XGBoost to avoid over-fitting.

For the prediction of traveler booking intent, we use a binary classification framework that applies a sigmoid function $\sigma(\cdot)$ to return a probability value. In booking value conditional on booking prediction, we use a regression framework that applies log transformation to improve accuracy.

\subsection{Skip-gram Sequence Model}
In order to generate feature learnings for traveler's session sequence to augment handcrafted features, we used a skip-gram model in our context attempts to predict listings $l_i$ before and after a given listing $l_{i-c}$ and $l_{i+c}$ viewed in a traveler session $\mathrm{S}_j$ , based on the premise that traveler's view of listings in the same session signals the similarity of those listings. We use a shallow neural network with one hidden layer with lower dimension for this purpose. The training objective is to find the listing local representation that specifies surrounding most similar manifold. More formally the objective function can be specified by the log probability maximization problem as follows:
\begin{equation}
\frac{1}{S}\sum_{s=1}^S\sum_{-c\leq i \leq c, i\neq 0} \log p(l_{i+j}|l_i)
\end{equation}
where $c$ is the window size representing listing context. The basic skip-gram formulation defines  $p(l_{i+j}|l_{i})$ using softmax function as follows:
\begin{equation}
p(l_{i+j}|l_{i}) = \frac{ \text{exp}(\nu_{l_{i+j}} ^T \nu _{l_{i}} ) }{\sum_{l=1}^L \text{exp}(\nu_{l} ^T \nu_{l_{i}} ) }
\end{equation}
where $ \nu_{l}$ and $\nu_{l_{i}}$ are input and output representation vector or neural network weights, and $L$ is the number of listings available on our platform. To simplify the task, we used the sigmoid formula, which makes the model a binary classifier, with negative samples, which we draw randomly from the list of all available listings on our platform\cite{mikolov2013distributed}. Formally, we use the following formula:
$p(l_{i+j}|l_{i}) = \frac{ exp(\nu_{l_{i+j}} ^T \nu _{l_{i}} ) }{1 + \text{exp}(\nu_{l_{i+j}} ^T \nu _{l_{i}} ) }$
for positive samples, and the following formula for negative ones:
$p(l_{i+j}|l^{0}_{i}) = \frac{1}{1 + \text{exp}(\nu_{l_{i+j}} ^T \nu _{l_{i}} ) }$.

We have two more issues to address, sparsity and heterogeneity in views per item. It is not uncommon to observe a long tail distribution of views for the listings. For this purpose we leverage approaches mentioned by \cite{mikolov2013distributed} 
wherein especially frequent items are downsampled
using the inverse square root of the frequency. Additionally we removed listings with very low frequency. 
To resolve the cold start issue, we leverage the contextual information that relates destinations (or search terms) to the listings based on the booking information. Formally, considering that the destinations $d_1, d_2, ..., d_D$ are driving $p_{id_1}, ..., p_{id_D}$, proportion of the demand for a given listing, we form the expectation of the latent representation for each location using $\nu_d=\frac{1}{N}\sum_{l=1}^Lp_{ld}\nu_{l}$, where $N$ is the normalizing factor. Then, given latitude and longitude of the cold listing (for which we have no data), we form the belief about the proportion of demand driven from each of the search terms $p_{jd_1}, ..., p_{jd_D}$. Then, we use our destination embedding from the previous step to find the expected listing embedding for the cold listing as follows $\nu_{l_j}=\sum_{d=1}^Dp_{jd}\nu_{d}$.

\subsection{Deep Average Network (DAN) and Alternatives}
\begin{figure}[htbp]
  \includegraphics[height=6cm, width=8cm]{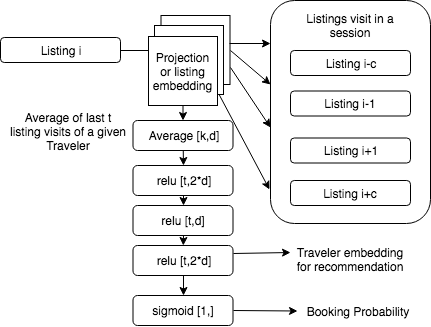}
  \caption{Deep Average Network (DAN) on the top of skip-gram network.}
  \label{fig:skipDAN}
\end{figure}
In the second stage, given the listing's embedding from the previous stage we model traveler embeddings using a sandwiched encoder-decoder non-linear Relu function. In contrast to relatively weak implicit view signals, in this stage we leverage strong booking signals as a target variable based on historical traveler listing interaction. We have various choices for this purpose including Deep Average Network with Auto-Encoder-Decoder, Long Short Term Memory(LSTM), and Attention Networks. The simplest approach is to take the point-wise average of embedding vector and use it directly in the model. The second approach could be to feed the average embedding into a dimensionality expansion and reduction non-linear encoder-decoder architecture, or Deep Average Network to extract the signals \cite{iyyer2015deep}. Hypothetically, this architecture may project the embeddings first into a larger space to isolate noise and then into smaller space to remove it\cite{broderick2005integrated}. The third approach could incorporate LSTM networks \cite{lang2017understanding} \cite{Sheil2018PredictingPI}. Hypothetically, this architecture may emulate the travelers' memory signal gathering and forgetting in the shopping funnel\cite{murray1991test}. The fourth approach could have an attention layer on top of the LSTM \cite{zhou2016attention}. Hypothetically, this architecture may capture an extra step in travelers' trip booking behavior in putting different weights on signals accumulated in their memory\cite{lynch1982memory}.

We take a probabilistic approach to model traveler booking events based on the embedding vectors of historical units they have interacted with. Formally, given the traveler embeddings (or last layer of the traveler booking prediction neural network), the probability of the booking is defined as:
\begin{equation}
\mathrm{P}(Y_j|\mathrm{S}_j, C_j) = \text{sigmoid}(f(\nu_{j.}))
\end{equation}

Then, the Deep Average Network layers are defined as:
\begin{align}
f(\nu_{j.}) &= \text{relu}(\omega_1 \cdot h_2(\nu_{j.}) + \beta_1)\\
h_1(\nu_{j.}) &= \text{relu}(\omega_2 \cdot h_1(\nu_{j.}) + \beta_2)\\
h_2(\nu_{j.}) &= \text{relu}(\omega_3 \cdot \frac{1}{k}\sum_{i=1}^{t}{\nu_{ji}}) + \beta_3)
\end{align}

Alternatively, we can use an LSTM network with forget, input, and output gates as follows:
\begin{multline}
f(\nu_{j}^{t}) = \text{sigmoid}(\omega_f [h_t, \nu_j^t]+\beta_f)
\cdot f(\nu_{j.}^{t-1})\\
+ \text{sigmoid}(\omega_i[h_t, \nu_j^t]+\beta_i)\cdot \text{tanh}(\omega_c[h_{t-1},\nu_j^t]+\beta_c)
\end{multline}

And finally, we can also use an attention network on the top of LSTM network as follows:
\begin{equation}
f(\nu_{j}) = \text{softmax}(\omega^T \cdot h_{T})\text{tanh}(h_{T})
\end{equation}
where $\omega_., \beta_.$ are weight and bias parameters to estimate and $h_t$ represents the hidden layer parameter or function to estimate.

Among these models, DAN is more consistent with Occam's razor, so it is more parsimonious, and faster to train. However, LSTM and Attention Networks on the top of it are more theoretically appealing. As a result, from the pragmatic stand point, for millions of listings and travelers DAN seems to be more appealing for deployment as depicted in Figure \ref{fig:skipDAN}. We use adaptive stochastic gradient descent method to train the binary cross entropy of these neural networks. 
\begin{table}[htbp]
\caption{Model Notations}
\begin{center}
\begin{adjustbox}{width=\columnwidth}
\begin{tabular}{ll}
\hline
\textbf{Variable} & \textbf{Description} \\
\hline
 $\mathrm{S}_j$&  discovery and booking session of the traveler \\
 $a_i$ &  activities of traveler (click, view, booking) \\
 $T$ & length of the session \\
$l_j$ & listing that the traveler has interacted with\\
$Y_j$ & indicator variable representing traveler booking \\
$C_j$ & traveler session context, or set of listings viewed \\
    &    before the current listing \\
$\mathrm{P}(Y_j|\mathrm{S}_j, C_j)$ & conditional booking probability for traveler $j$ \\
$\mathrm{P}(V_j|\mathrm{S}_j, C_j)$ & conditional booking value probability for traveler $j$ \\
$\mathcal{U}$ & bid utility \\
$\mathcal{R}$ & probability of conversion \\
$\mathcal{M}$ & marginal value of conversion \\
$x_j$ & set of features \\
$y_j$ & set of labels \\
$\mathbb{F}(.)$ & additive function \\
$l$ & the loss function (log likelihood) \\
${L}^{(t)}$ & regularized objective function \\
$f_t(\mathbf{x}_j)$ & base learner function \\
$l_{i-c}$ & listing viewed in the same session with listing $i$ \\
& but is $c$ listings before it \\
$\nu_{l}$ & neural network weights \\
$d_i$ & a destination \\
$p_{id_1}$ &  proportion of the demand for a given listing \\
    & driven by destination $d_1$ \\
$\text{relu}$ & the rectifier function, or an activation function \\
& defined as the positive part of its argument\\
$\omega_i$ & weights of neural network \\
$\beta_i$ & biases or intercepts of neural network \\
$h_i(.)$ & hidden layers in the neural network \\
$f(.)$ & output layer of the neural network \\
$\text{sigmoid}$ & S shaped logistic function \\
$\text{tanh}$ & hyperbolic function \\
$t$ & representing epoch or time \\
$\omega_f$ & represents forgetting weights in the LSTM neural network \\
$\omega_i$ & represents input weights in the LSTM neural network \\
$\omega_c$ & represents new candidate weight in the LSTM neural network \\
$\text{softmax}$ & normalized exponential function \\
$[.,.]$ & concatenate vectors \\
\hline
\end{tabular}
\label{tab:Notation}
\end{adjustbox}
\end{center}
\end{table}
Table \ref{tab:Notation} summarizes the notation used in this section. In the next section we review how to deploy this model end to end.


\section{System Overview}

\begin{figure}[htbp]
  \includegraphics[height=6cm, width=9cm]{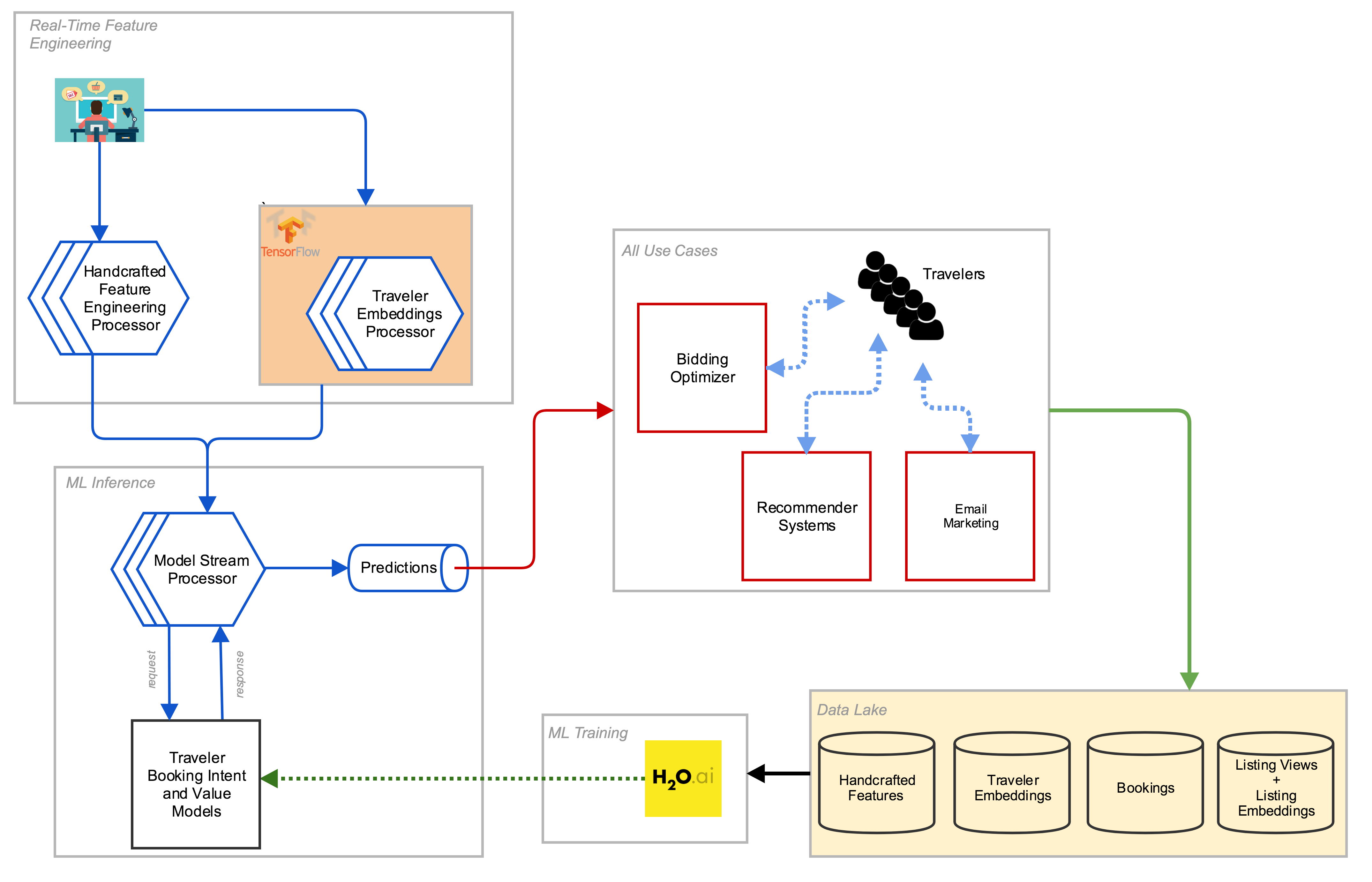}
  \caption{High-level overview of the Deep Personalized Re-targeting System. The solid blue arrows represent the asynchronous real-time flow of data during prediction time, the solid black arrow represents batch training data, the dashed green arrow depicts the loading of models from model repository, and the red solid red arrow shows the usage of predictions from in-house systems. Finally, the dashed light-blue lines represent interaction between travelers and in-house models, and the solid green arrow shows the logging of data related to the latter interaction to the Data Lake.}
  \label{fig:architecture}
\end{figure}

The high-level real-time system architecture is shown in Figure \ref{fig:architecture}. Each traveler's interaction (listing view, dated search, etc) leads to an event that is transformed into handcrafted features and if there is a listing view, traveler embeddings.  The Model Stream Processor, which has loaded the Traveler Booking Intent Classification and Booking Value Regression models, consumes the handcrafted features and traveler embeddings for each traveler, and calls the prediction functions of the two models. Predictions are consumed by the Bidding Optimizer. More specifically, we use thresholds to assign travelers to buckets $\left\{b_1, b_2,..., b_N\right\}$, where $N$ is the total number of buckets based on their booking probability and value predictions. We tune the thresholds to evenly distribute the density for each of the buckets. As a result, the expected average booking Revenue Per Click (RPC) in the buckets is monotonically increasing by design. From an empirical stand point, these buckets mirror the shopping funnel of the travelers, i.e. bucket $b_1$ contains travelers with the lowest predicted booking probability and values, resembling discovery stage, whereas $b_N$ contains travelers that are close to the end of the booking funnel. The Bidding Optimizer generates higher ROI with the same budget when these buckets with traveler identifiers are given to it. This happens as the Bidding Optimizer is guided to allocate financial resources to each bucket accordingly. 

For training, real-time handcrafted features and traveler embeddings are persisted in the S3\footnote{an Amazon Web Services cloud solution that provides object storage} based Data Lake where they are joined with booking data to get labels. Then, the XGBoost Traveler Booking Intent and Value models are trained on this data using H2O \cite{H2O}. Another offline process trains the Tensorflow \cite{tensorflow2015-whitepaper} Deep Average Network based traveler embeddings frequently using historical data from that Data Lake to address the seasonality of the traveler industry and cold start issue. Finally, traveler booking probabilities, values and embeddings are consumed by other systems and stakeholders apart from the bidding optimizer. Such systems are in-house recommender systems and email marketing related models. 

\section{Experiments and Results}

We compare the Traveler Booking Intent (TBI) accuracy-uplift of our Deep Average Network based approach to various baselines in this section. For offline evaluation, we merged the handcrafted features and the traveler embeddings, generated by all different model settings, and fed them to the TBI model.

\subsection{Methodologies}
In this subsection, we describe three baseline methods that we compare against our proposed Deep Average Network (DAN) on the top of Skip-Gram:
\begin{enumerate}
  \item \textbf{Random}: a heuristic rule that chooses a random listing embedding, among those listings a traveler has previously interacted with in the current session.
  \item \textbf{Averaging Embeddings}: a simple point-wise averaging of listing embeddings a traveler has previously interacted with, in the current session.
  \item \textbf{LSTM with Attention}: A recurrent neural network, inspired by \cite{lang2017understanding}, \cite{10.1007/978-3-030-10997-4_9} and \cite{Sheil2018PredictingPI}, that uses LSTM units and an attention mechanism on top of it in order to combine embeddings of listings a user has previously interacted with in the current session.
\end{enumerate}

\subsection{Datasets}
For the experiments, anonymized clickstream data was collected for millions of users from two different seven-day periods. The first dataset was used to generate embeddings using Deep Average Network and the LSTM with Attention. The second dataset was used to evaluate the learned embeddings on the Traveler Booking Intent Model. 

\subsection{Results}
We evaluated the performance of the Traveler Booking Intent model on the different settings using AUC, Precision, Recall and F1 scores. The best results of each model are shown in Table \ref{tab:comparison}. It shows that our proposed Deep Average Network approach contributes more uplift to the TBI model.

\begin{table}[htbp]
\caption{Comparison between Model Settings}
\begin{center}
\begin{adjustbox}{width=\columnwidth}
\begin{tabular}{|c|c|c|c|c|}
\hline
\textbf{Algorithm} & \textbf{AUC}& \textbf{Precision}& \textbf{Recall}& \textbf{F1-Score} \\
\hline
Random & 0.973 & 0.821 & \textbf{0.633} & 0.715  \\
Averaging Embeddings & 0.971 & 0.816 & 0.628 & 0.71 \\
LSTM + Attention & 0.976 & 0.877 & 0.62 & 0.727 \\
\textbf{DAN} &  \textbf{0.978} & \textbf{0.888} & 0.628 & \textbf{0.735} \\
\hline
\end{tabular}
\label{tab:comparison}
\end{adjustbox}
\end{center}
\end{table}

Moreover, Table \ref{tab:uplift} shows the TBI performance improvement when the DAN generated traveler embeddings are merged with the initial handcrafted features. Our finding suggests embeddings have comparative predictive power to handcrafted features.

\begin{table}[htbp]
\caption{Performance Uplift to TBI Model}
\begin{center}
\begin{adjustbox}{width=\columnwidth}
\begin{tabular}{|c|c|c|c|c|}
\hline
\textbf{Settings} & \textbf{AUC}& \textbf{Precision}& \textbf{Recall}& \textbf{F1-Score} \\
\hline
Only handcrafted Feat. & 0.975 & 0.817 & \textbf{0.651} & 0.724 \\
\textbf{Handcrafted + DAN Feat.} &  \textbf{0.978} & \textbf{0.888} & 0.628 & \textbf{0.735} \\
\hline
\end{tabular}
\label{tab:uplift}
\end{adjustbox}
\end{center}
\end{table}

\section{Conclusion}
In this paper, we introduced a hybrid deep learning framework for a massive vacation rental marketplace. Deployed in an end-to-end manner, this pragmatic framework aims to help solve challenges in prediction of traveler shopping journey within a re-targeting scope. Our results show that by leveraging neural network traveler embeddings trained on session logs we are able to enhance the prediction of our original booking intent model which used handcrafted features. We also find that there is a pragmatic sweet spot between expensive complex deep neural networks and simple shallow neural networks that can increase the performance of the boosting tree model, based on offline analysis. Furthermore, incremental complexity in this model enables extracting traveler embeddings, which can also be used for personalizing recommender systems. To further improve and extend this work, we are starting to explore options to transfer the learnings to other problems in the marketing domain which are subject to data sparsity issues. It can also be beneficial to infuse other contextual spatio-temporal information into our model to help drive a smooth and personalized full-cycle traveler booking experience.

\section{Acknowledgments}
This project is a collaborative effort between the recommendation, marketing data science and growth marketing teams. The authors would like to thank Chandri Krishnan, Andrew Reuben, Travis Brady, Wenjun Ke, Ali Miraftab and Ravi Divvela for their contribution to this paper.

\bibliographystyle{IEEEtran}
\bibliography{IEEEabrv,IEEEexample}


\end{document}